# A Modified CSMA/CA Protocol for OFDM Underwater Networks: Cross Layer Design

Yanling Yin[†]
National Laboratory of Underwater Acoustic Technology
Harbin Engineering University
Harbin, Heilongjiang, China
yyling@uw.edu

Sumit Roy
Dept. of Electrical Engineering
University of Washington
Seattle, WA, USA
sroy@u.washington.edu

Payman Arabshahi
Applied Physics Laboratory and Dept. of Electrical Engineering
University of Washington
Seattle, WA, USA
paymana@ u.washington.edu

## ABSTRACT
The underwater acoustic channel continues to present significant challenges to efficient throughput performance of underwater acoustic sensor networks (UASNs) in varying scenarios. As a result, cross-layer approaches that explore joint PHY/MAC strategies are worthy of further exploration. We consider a recent high-speed OFDM modem and propose a new cross-layer solution based on modified CSMA/CA, for a canonical star network topology with few nodes (the most common scenario in UASNs). Some innovations to an adaptive OFDM PHY link are developed to jointly select the modulation, convolutional coding and frequency diversity order (different transmission modes) for matching varying channel conditions. Additionally, receiver logic that disambiguates the cause of packet loss between a) that caused by channel vs. b) that due to collisions is used to modify the ARQ/backoff logic for retransmissions with CSMA/CA random access. Simulation results reveal that the cross-layer design can effectively increase network throughput.

## Categories and Subject Descriptors
C.2.1 [**Network Architecture and Design**]: Centralized networks, Wireless communication; C.4 [**PERFORMANCE OF SYSTEMS**]: Design studies.

## 1. INTRODUCTION
The growing interest in underwater acoustic sensor networks (UASNs) reflects their increasing use in a variety of environments for commercial exploitation and scientific exploration such as climate monitoring, pollution detection, and oceanographic data collection. In order to meet the variety of applications, robust and flexible UASNs are needed that provide the right design trade-offs in terms of throughput, end-to-end delay and error rates while reducing nodal energy consumption. However, due to the complex and variable underwater acoustic channels – characterized by high attenuation, strong interference and large Doppler, the performance of UASNs is extremely constrained [1]. Underwater modems today achieve low data rate (kbps) at typical desired ranges (kms) and often suffer from high link error rate. The long propagation delay seriously reduces the effectiveness of key MAC protocol components (such as carrier sensing for collision avoidance) and also compromises how the network reacts to adverse conditions [2][3]. Also, the energy consumption in UASNs continues to be a challenge as nodes are usually battery powered.

Recently, high data rate OFDM PHY modems such as [4] have become available; but their potential in underwater scenarios for high throughput operations remains unrealized. Cross-layer design provides a pathway to achieving higher network throughput as well as energy utilization via jointly optimizing the functionalities at different layers of the protocol stack. This paper focuses on the design aspects of the interaction between the PHY and MAC layer in response to the unknown and varying underwater channel. In order to simplify the problem, we adopt a simple star topology to analyze the performance of the cross-layer design. Our design rests on adaptive OFDM modulation which selects the modulation, coding and frequency diversity to achieve variable rate transmission. We add to this a modified CSMA/CA mechanism driven by information from the PHY (hence cross-layer) to improve the MAC throughput for a star topology.

The rest of the paper is organized as follows. In Section 2, we briefly review the cross-layer design in UASNs. In Section 3, we describe the proposed PHY/MAC cross-layer design including the modified MAC protocol, adaptive link layer communication and the interaction between PHY and MAC layers. Section 4 examines the performance of the cross-layer design through simulation. Section 5 concludes this paper.

## 2. RELATED WORK
The main objective of cross-layer design is either saving energy (thus prolonging the lifetime of the network) or maximizing the network end-to-end throughput. In [5], the authors propose an extension of the RMAC [6] protocol with transmission power control (RMAC-PC), to improve the energy efficiency of PMAC protocol using MAC-PHY cross-layer optimization. The basic idea is to utilize the knowledge of latencies and use the optimum power for transmission. In [7], power and frequency allocation are adjusted as a practical means of optimizing the overall performance across the physical, MAC and routing layers in a 3 dimensional multi-hop network. The distance-aware collision avoidance protocol (DACAP) and focused beaming routing (FBR) are adopted as the MAC and routing protocols. The average energy per bit consumption is reduced by adjusting the power, center frequency, and bandwidth in accordance with the network node density. In [8], the authors propose distributed routing algorithms for UASNs with MIMO-OFDM links. They attempt to reduce energy consumption in a cross-layer fashion by jointly

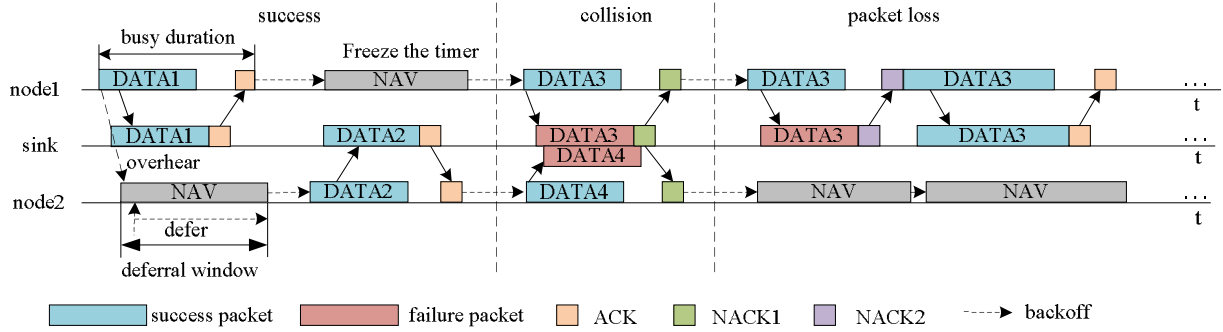

**Figure 1. Timeline flowchart of the modified CSMA/CA protocol.**

i) selecting its next hop, ii) choosing a suitable transmission mode, and iii) assigning optimal transmit power on different subcarriers.

Much of the literature focuses on minimizing the energy consumption or the routing layer design [9] in multi-hop UASNs. In contrast, in this work we focus on a single hop star network and present a cross-layer solution on top of the OFDM PHY layer. We focus on the interaction and information exchange across these two layers with the aim of maximizing the throughput with a view to future *software defined* underwater modems that will translate these ideas into the field.

## 3. PHY/MAC CROSS-LAYER PROTOCOL

The key idea of the cross-layer solution is that the MAC layer uses information from the PHY layer to disambiguate the cause of packet loss and take appropriate action. The PHY layer provides information such as estimated SNR to help the MAC layer modify the ARQ/backoff logic for subsequent retransmission. Also, the PHY layer jointly adapts modulation, forward error correction (FEC) and frequency diversity functionalities to varying channel conditions and to achieve a stable packet error rate (PER).

### 3.1 Modified CSMA/CA Protocol

Although ALOHA based protocols exploit the full bandwidth of the system, frequent collision leads to a low throughput and high energy consumption, due to lack of coordination. The logical next step of including carrier sensing in multiple access (CSMA) to mitigate collisions is hampered by the long propagation delay that reduces channel utilization dramatically; see [10]. We therefore propose a variation of unslotted CSMA/CA (the baseline 802.11 access protocol) as outlined below.

In the proposed modified CSMA/CA protocol, we do not use the RTS and CTS frames to cut down on protocol overhead. The node senses the channel by overhearing the packet (this is accomplished by detection of the preamble, see Fig.2(a) and *virtual* carrier sensing (instead of *physical* carrier sensing) by decoding the header) that includes the network allocation vector (NAV) which informs all nodes as to how long the current transmission will occupy the channel. The receiver is always open and keeps correlating the input signal with the known preamble. If the correlation peak exceeds a pre-set threshold, it indicates the start of packet and the node proceeds to decode the header that includes the recipient ID. If it is the intended receiver, it then decodes the payload. Otherwise, it drops the packet and sets its own NAV to the busy duration, see Fig.1. Note that the busy duration $T_{busy}$ is defined as

$$T_{busy} = T_{data} + T_{ACK} + 2T_{delay} + T_{other} \qquad (1)$$

where the data packet duration $T_{data}$ is variable for different data rates. $T_{ACK}$ denotes the ACK packet duration; $T_{delay}$ is the (one-way) propagation delay between the source and destination node; and $T_{other}$ is the overhead such as processing time. In Eq. (1), $T_{data}$, $T_{ACK}$ and $T_{other}$ are generally known a-priori or can be reasonably estimated, but $T_{delay}$ needs to be measured for the specific scenario. This can be done during initialization by a one-time short control message exchange to measure the propagation delay from each source to sink. Due to node drift, the propagation delay can be expected to change slowly and hence it needs to be updated after each first time successful transmission.

The deferral interval is set equal to $T_{busy}$ where upon every node picks a random backoff value initially in [0, $CW_{min}$], sets a backoff counter and begins countdown. After a successful transmission, the node randomly backsoff and the backoff window is randomly chosen from 0 to $CW_{min}$. In the event of a collision, the backoff window is doubled at each stage (up to a maximum) per the binary exponential backoff algorithm. An optimal value for the initial contention window ($CW_{min}$) is chosen for different values of the network nodes according to [11]. During the backoff period, if any node overhears/detects a packet on the channel, it freezes the timer for the NAV duration as stated previously (see Fig.1). When a node's countdown timer expires, it transmits its packet. A collision may occur in the event that two (or more) nodes' timers expire simultaneously. Once the node transmits a packet, it waits till it receives an ACK/NACK or encounters a timeout. The subsequent steps in MAC transmitter state transition (in response to the above events) are explained in detail in Sec. 3.2 and shown in Fig. 2.

### 3.2 Cross-layer Design

In our cross layer design, we introduce different NACK packets in case of packet decode failure at the receiver to indicate the cause (to the sender). We classify the cause of packet loss either due to a) single transmission encountering a bad channel or b) two (or more) packets overlapping at the receiver (collision). The receiver sends a NACK to inform the sender of either case. If the packet loss is caused by a bad channel, the sender immediately retransmits the packet using a lower data rate without any backoff. If the packet loss is caused by collision, the sender backs off in time and retransmits the packet.

A flowchart for the proposed cross-layer solution is shown in Fig.2(b). There are four states/actions for the receiver:

I) If the received packet is successfully decoded, the receiver sends an ACK to the sender. Meanwhile, the physical layer estimates the effective SNR (ESNR) that is fed back in the ACK and is used for mode selection for the next transmission. ESNR

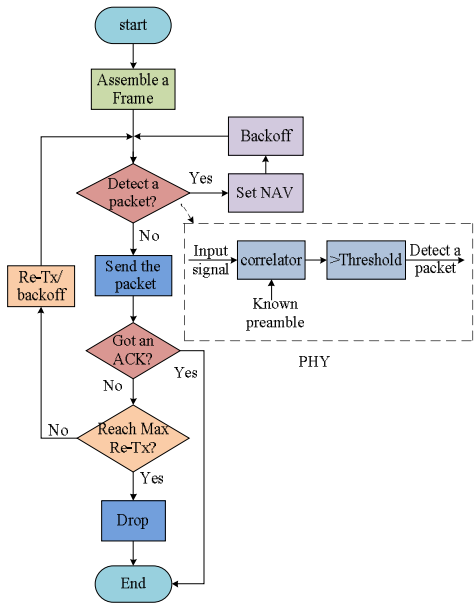
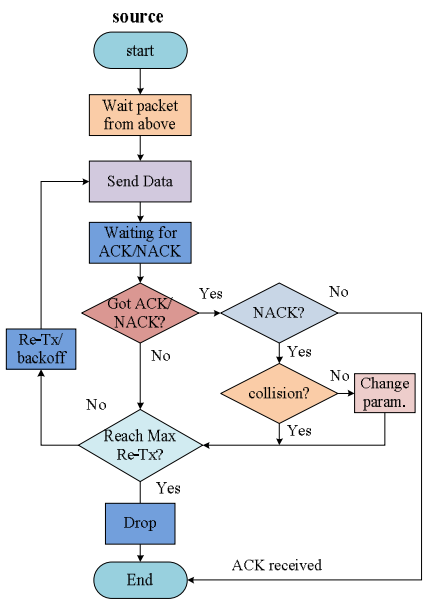

**Figure2(a). Flowchart of modified CSMA/CA protocol.**  **Figure2(b). Flowchart of the proposed cross-layer solution.**

computation and transmission mode selection is discussed in Section 3.3.2.

II) If the receiver detects a single preamble but fails to decode the payload caused by the bad channel, it sends a NACK1 (type1) to the sender to indicate that packet loss is due to channel conditions.

We next consider cases where two or more packets overlap at the receiver ('collision'). In our simulation experience, we find that in such collision scenarios, *all* overlapping packets are lost (i.e. decode failure). However, we distinguish between two scenarios of overlap: a) where the latter packet is sufficiently delayed and overlaps the payload of the first, thereby allowing the receiver to detect both preambles and declare a collision and send that information back to the senders (broadcast NACK2), and b) where the two packets overlap substantially resulting in preamble detect failure, and hence missed detection (In the (unlikely) scenario that the 1st packet is detected and correctly decoded, we will modify our MAC to send a unicast ACK to the 1st sender, while the other sender times out.)

III) If two or more packets are partially overlapped such that the receiver detects collision (see Sec. 3.3.3), it broadcasts a NACK2 (type 2).

IV) If two or more packets are received near synchronously or the first preamble cannot be detected by the receiver for any reasons, this results in missed detection at the receiver and subsequent timeout at the sender.

On the sender side, after transmitting a packet, it waits for an ACK/NACK or timeout. If the sender

I) receives an ACK, it prepares for the next packet.

II) receives a NACK1, it decreases the data rate (selects a lower transmission mode, see section 3.3) and retransmits the packet immediately.

III) receives a NACK2 or encounters a timeout, it backs off and retransmits the packet using the same transmission mode.

After three retrials, if the packet still fails successively, the sender drops the packet.

### 3.3 Adaptive OFDM communication

Recently, OFDM has been introduced for underwater acoustic communications to achieve higher data rates (e.g. in the AquaSeNT modem [4]). Our work is therefore based on an OFDM enhanced cross-layer solution to improve adaptation to the channel and hence achieve higher throughput. Prior work in underwater OFDM includes [12], which proposes two methods for adaptive OFDM modulation over time-varying channels. The system adjusts the modulation level or both the modulation level and transmit power to maximize the system throughput under a target average bit error rate (BER). The optimal modulation level and power are decided by the feedback predicted channel. In this scheme, the sender needs to know the channel state information and requires the channel to change slowly. In [13], an adaptive modulation and coding (AMC) scheme is proposed for underwater acoustic OFDM system. In this scheme, LDPC coding is adopted and the system is constructed with a finite number of transmission modes. There are four hydrophones at the receiver to increase the received SNR. The AMC system is also considered in the single carrier modulation system [14] and Turbo coding is used in [15]. In this paper, modulation, FEC (convolutional coding) and frequency diversity are jointly selected to achieve a target PER over time-varying underwater acoustic channels.

#### 3.3.1 Transmission mode selection

We select different parameters to form a group of transmission modes. BPSK, QPSK and 8PSK are provided as the constellation size. The convolutional code is adopted and the candidate coding rates are 1/4, 1/3, and 1/2. Distinct from traditional AMC system, frequency diversity is also considered in our design. Due to wind, shipping noise and other human activities, the spectrum of the fading channel is usually not flat in lake or shallow water [12]. Additionally, due to the complex geography of bottom and the environment, only decreasing the modulation level and the coding rate (e.g. BPSK and 1/2 coding rate) cannot achieve the desired performance. Frequency diversity is a good choice to combat the frequency selective channel [16]. Therefore in our design, we jointly adapt the modulation level, coding rate and frequency diversity order for matching varying channel conditions.

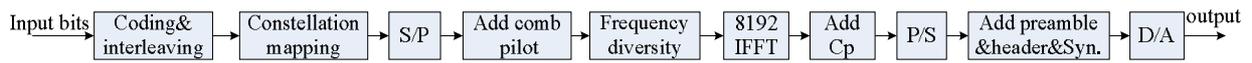

**Figure 4. Transmitter diagram of the OFDM system.**

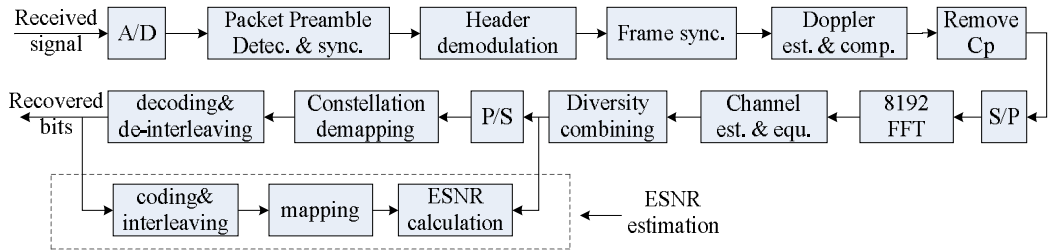

**Figure 5. Receiver diagram of the OFDM system.**

These three parameters represent many possible combinations or modes and the transmitter attempts to choose the one which yields the best performance. Table 1 shows the selected parameters for different transmission modes.

**Table 1. Transmission modes and parameters.**

| Modes | Modulation | Diversity order | Coding rate | Data rate (kbps) |
|---|---|---|---|---|
| mode 1 | BPSK | 3 | 1/2 | 0.658 |
| mode 2 | QPSK | 3 | 1/2 | 1.317 |
| mode 3 | QPSK | 1 | 1/4 | 1.984 |
| mode 4 | QPSK | 1 | 1/3 | 2.645 |
| mode 5 | QPSK | 1 | 1/2 | 3.967 |
| mode 6 | 8PSK | 1 | 1/2 | 5.950 |

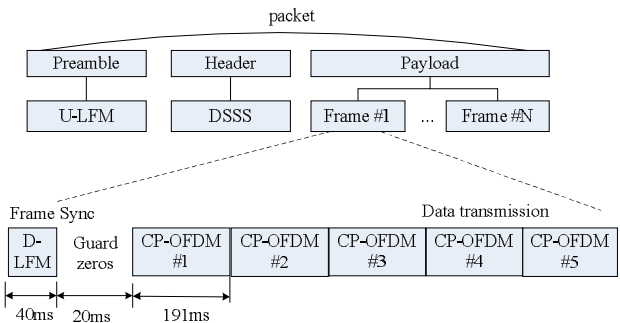

**Figure 3. Data packet structure and frame structure.**

The data packet structure is shown in Fig.3. Different from packet structure in 802.11 protocols, each part of packet is specially designed for underwater acoustic communication. The data packet contains three parts: the first part is the preamble which enables the receiver to detect and synchronize the incoming data packet. Here, a linear up-chirp (U-LFM) pulse signal is adopted as the preamble which is simple to be detected and robust over the multipath and Doppler distorted channels. The second part is the header. It contains the demodulation information of the packet including the source node ID, the destination node ID, the packet length, the transmission mode, etc. The header needs to be transmitted reliably and we choose DSSS modulation for robust decoding in typical underwater channels. The last part is the payload that contains the net data needed from the source node to the destination node. The payload is transmitted by OFDM modulation and contains several frames decided by the packet size.

The frame structure is also shown in Fig.3. At the beginning of the frame is a synchronization signal. In order to be different from the packet preamble, the synchronization signal can use a linear down-chirp (D-LFM) signal. It is followed by the cyclic prefix padded OFDM (CP-OFDM) block. There is a guard interval between the synchronization and the CP-OFDM blocks. This guard interval is longer than the channel delay spread. One frame contains five OFDM blocks.

The transmitter and receiver diagrams of the system are shown in Figs. 4 and 5 respectively. Note that the transmitter implements frequency diversity as follows. We divide the whole bandwidth into $M$ non-overlapping segments, where $M$ is the diversity order. The mapped data stream, after serial-to-parallel (S/P) conversion, is repeated on the data subcarriers of each diversity channel and the pilots are uniformly distributed at a regular interval. Then an 8,192 points IFFT is implemented and a cyclic prefix is added. The OFDM blocks, after parallel-to-serial (P/S) conversion, are added the preamble, header and the frame synchronization, to form a data packet. The packet is then transmitted into the channel.

On the receiver side, the receiver keeps sampling the data and the sampled data pass through the following steps:

1. Preamble detection and synchronization: detect the preamble and synchronize the packet using a cross-correlator.
2. Header demodulation: demodulate the header, get the packet's information, such as source ID, destination ID, packet length, etc.
3. Frame synchronization: synchronize the frame and process the data block by block.
4. Doppler estimation and compensation: before the FFT process, Doppler should be compensated first. Doppler can be estimated by the method in [17] using LFM signals and compensated by re-sampling. In the simulation, we assume that the Doppler can be perfectly removed.
5. Remove cyclic prefix; perform S/P and 8,192 FFT: remove the cyclic prefix (Cp); covert the data stream from serial to parallel; implement 8,192 points FFT.
6. Channel estimation and equalization: estimate the channel and equalize the data. Least square (LS) channel estimation is used.
7. Diversity combining: combine each frequency diversity. Maximal-ratio combing (MRC) is used for diversity combination.
8. P/S, constellation demapping, decoding and de-interleaving: covert the data stream from parallel to serial, demap the data, decode and de-interleave the data, get the recovered bits.
9. ESNR estimation: ESNR is a performance metric for mode selection (see below). After we obtain the recovered bits,

recode, re-interleave and re-map the bits, we use the mapped data and the recovered data in frequency domain after MRC to estimate the ESNR.

Fig. 6 shows the packet error rate (PER) performance for the 6 transmission modes. The simulation parameters are shown in Table 2. The multipath channel used in the simulation is generated by BELLHOP [18] for the sound speed profile in Fig.7. In the simulation, the center frequency is 9kHz, and the transmitter and the receiver depths are 10m. The distance between the transmitter and the receiver is 1km. The simulated channel impulse response is shown in Fig.8. This channel is also used in the following simulations.

**Table 2. OFDM parameters.**

| Packet size | 400bytes | Subcarrier spacing | 5.86Hz |
|---|---|---|---|
| Sampling rate | 48kHz | Number of subcarriers | 1,025 |
| Center frequency | 9kHz | Number of data carriers | 768 |
| Signal bandwidth | 6kHz | Number of pilot carriers | 257 |
| Symbol duration | 170.7ms | Constellation size | 2, 4, 8 |
| Cp duration | 20ms | Diversity order | 1, 2, 3 |
| FFT/IFFT | 8,192 | Coding rate | 1/4, 1/3, 1/2 |

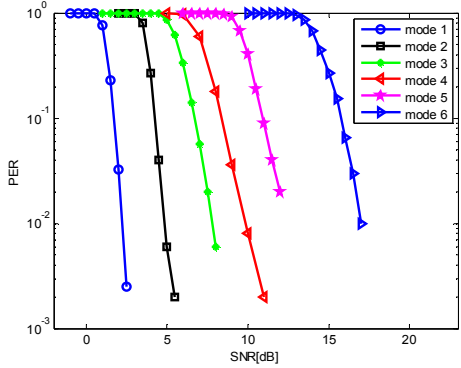

**Figure 6. PER of 6 transmission modes.**

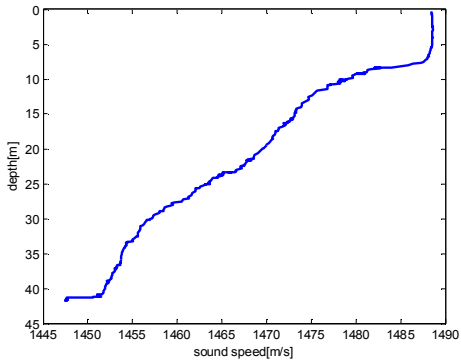

**Figure 7. Sound speed profile.**

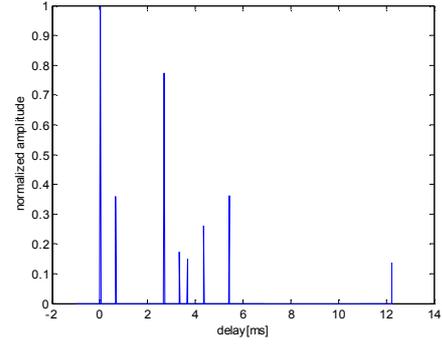

**Figure 8. Simulated channel impulse response.**

### 3.3.2 Performance metric

Since the sender selects the transmission modes, another critical issue is to find a performance metric for mode switching. Input SNR (ISNR) is traditionally used and can be easily calculated in the time domain at the receiver by comparing the received signal to the noise power on a per subcarrier basis. However, ISNR may not accurately reflect the performance of OFDM in underwater channels since it does not capture the impact of multipath and Doppler effects. Pilot SNR (PSNR), proposed in [19], more accurately captures the channel imperfection as it estimates the SNR by measuring the received signal power at pilot and null subcarriers. However, it is computed prior to the channel estimation which ignores the impact of the channel estimation to the system performance. It therefore might not be a consistent performance metric. In [13] the authors propose the effective SNR (ESNR) as the performance metric. The ESNR is computed after the receiver has successfully decoded a message. In this calculation, the noise item contains not only the ambient noise, the residual inter-carrier interference, but also the noise due to the channel estimation error. The ESNR can be calculated as follows,

$$ESNR = \frac{E_{k \in S_D}\left[\left|\hat{H}[k]s[k]\right|^2\right]}{E_{k \in S_D}\left[\left|z[k]-\hat{H}[k]s[k]\right|^2\right]} \quad (2)$$

where $S_D$ is the ensemble of the data subcarrier; $\hat{H}[k]$ is the estimated channel response of the $k$-th subcarrier in frequency domain; $z[k]$ is the frequency observation at subcarrier $k$; and $s[k]$ is the transmitted symbols on subcarrier $k$ which can be known since it is done after successful decoding of the message. Simulation and experimental results in [13] show that ESNR has consistent performance over various channels. We therefore adopt ESNR as the performance metric to select a suitable transmission mode.

### 3.3.3 Collision detection

In our cross-layer design, the PHY layer needs to disambiguate the packet loss between those caused by channel or those due to collisions; this information is used to modify the ARQ/backoff logic for retransmission. If the receiver can detect the preamble but fails to then decode the payload, the packet loss is caused by a bad channel. The preamble (up-chirp) can be expressed as

$$s(t) = \cos\left(2\pi ft + \pi(B/T)t^2\right) \quad (3)$$

where $f$ is the start frequency, $B$ is the bandwidth, and $T$ is the time duration. In our simulation, the sampling rate is 48kHz. LFM lasts for 40ms and the bandwidth is from 6kHz-12kHz. The collision detection algorithm can be explained using the pseudo-code in Table 3. The receiver keeps sampling the received signal and correlating the incoming data with the locally known preamble. If the correlation result exceeds the threshold, it indicates a packet arrival. The receiver records this time and begins decoding the header (to obtain the payload duration information) and then the payload. When decoding the first packet, the first packet payload, the receiver continues the preamble correlation. If another preamble is detected, the receiver records the arrival time of the second packet and calculates the duration between two arrival instants. If this duration is shorter than the first packet duration, it means that there is a collision.

**Table 3. Pseudo-code of collision detection algorithm.**

1: Correlate the received signal with the local preamble
2: **If** the correlation result > threshold **then**
    Record the peak position, decode the header, obtain the packet duration
    **else**, continue with the correlation.
3. **If** another preamble is detected and the duration of two peak positions<the data packet duration
→ collision detected

The collision detection probability for a single packet depends on the post-correlation SNR at the receiver for the preamble. In the case of two overlapping packets where the latter packet arrives during the payload of the first, the first packet has a higher detection probability than that of the latter packet. First, let us look at the detection probability of the LFM signal without overlap over AWGN and multipath channel shown in Fig.9. There is a small drop in multipath channel compared to that in AWGN channel. Because of multipath distortion, the correlator is not a matched filter. If we wish to realize the matched filter, channel estimation is needed before transmitting the packet, which is an extra expenditure. The detection probability is high enough in our simulation scenario that we just correlate the received preamble with the transmitted one. From Fig.9 we can see that, in the multipath channel, it requires about -9dB to keep the detection probability higher than 90%.

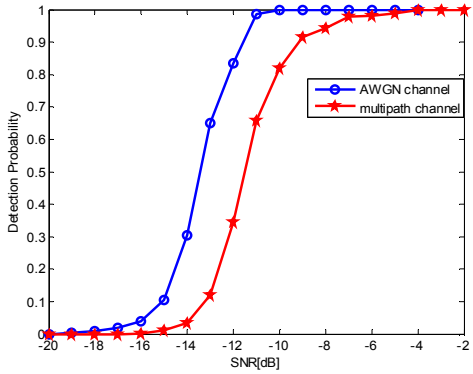

**Figure 9. LFM detection probability without overlap.**

When two packets overlap, the detection probability of the overlapped LFM signal is shown in Fig.10. When the preamble of the 2$^{nd}$ packet overlaps with the OFDM symbol of the 1$^{st}$ packet, we can observe that even when the interfering packet is much stronger than the desired signal (low SIR), the receiver still has a high detection probability.

Now, let us analyze the detection probability to see whether it meets system requirements. Assume the transmit power (Pa) of the node is 2 W; we express the source level (SL) as

$$SL=10\log_{10}(Pa)+170.77=173.77 dBre\ 1\mu Pa \quad (4)$$

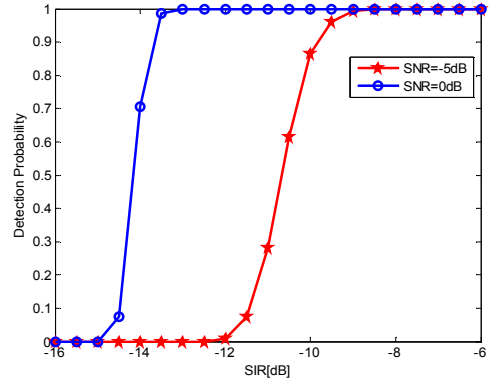

**Figure 10. LFM detection probability with 2 packets overlap.**

For a point source, the wave propagation obeys spherical spreading, so for 1,000m distance (used in Section 4), the spreading loss can be calculated as

$$TL_1=20\log_{10}(r)=20\log_{10}(1000)=60dB \quad (5)$$

According to Thorp's empirical formula [20], the absorption coefficient (decibels per kilometer) is

$$10\log(a(f))=\frac{0.11f^2}{1+f^2}+\frac{44f^2}{4100+f^2}+2.75\times10^{-4}f^2+0.003 \quad (6)$$

where $f$ is in kHz. For a center frequency of 9kHz, the absorption loss is $TL_2$=8.43dB. The net transmission loss (TL) is then

$$TL=TL_1+TL_2=68.43dB \quad (7)$$

For the case when transmitter and receiver are ormidirectional, the transmitted and received directivity indices are zero. Consider the near shore shallow water with heavy ship traffic, assume the noise level (NL) is between 80 and 100dB (6kHz bandwidth). According to the passive sonar equation, the received SNR can be estimated as

$$5.34dB<SNR=SL-TL-NL<25.34dB \quad (8)$$

Since the minimum SNR is 5.34dB, the proposed collision detection algorithm works reliably in our simulation scenario.

## 4. SIMULATION RESULTS

To assess the performance of the cross-layer design, we have used MATLAB for simulation. A star topology network is considered, composed of a varying number of active nodes, randomly located over a 1,000m by 1,000m square region; all nodes are fully connected. The sink node is located in the center of the region. The packets generated by each node follow the Poisson distribution. The total number of packets offered per second is kept the same, even as the number of nodes is varied. The multipath channel used in the simulation is shown in Fig.8.

In the simulation, the payload of each data packet contains 400 bytes. The data rate is 658bps (mode 1). The data packet and ACK last for 5.36s and 0.5s respectively. The maximum propagation delay from source node to the sink is 0.47ms. Figure 11 shows the throughput of different number of nodes vs. the packet arrival rate (packets/s). From the figure we can see that with the packet arrival rate increasing, the normalized throughput also increases and then stays flat. There is a little drop in the maximum throughput as the number of nodes increases. When the number of nodes is large, the maximum throughput reaches a stable value. The maximum

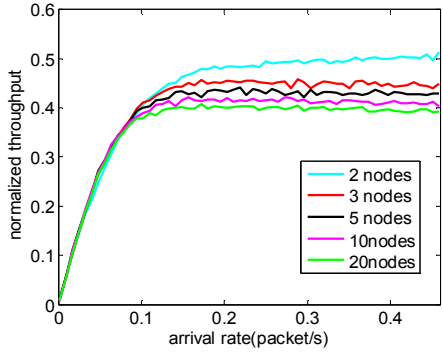

Figure 11. Normalized throughput of different numbers of nodes.

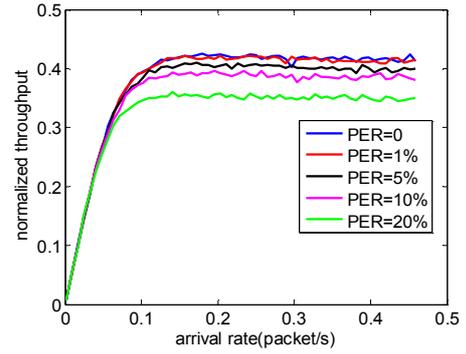

Figure 13. Normalized throughput with different PERs.

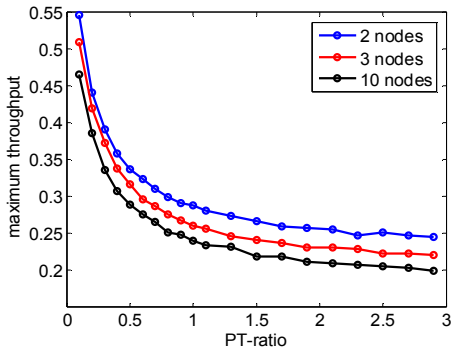

Figure 12. Maximum throughput vs. PT ratio.

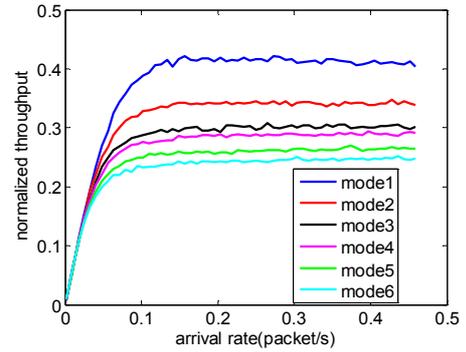

Figure 14. Normalized throughput of different modes.

throughput is related to the data packet duration and the propagation delay.

Figure 12 shows the maximum throughput vs. PT-ratio with different number of nodes. "PT-ratio" in defined as follows [21]

$$\text{PT-ratio} = \frac{\text{Average propagation delay (P)}}{\text{Packet transmission time (T)}} \quad (9)$$

When the propagation delay is smaller than the packet duration, the maximum throughput can achieve a good throughput. With the propagation delay increasing, the maximum throughput drops off and then achieves a stable value.

Figure 13 shows the throughput with different PERs. From the figure, we can see that with the PER increasing, the throughput decreases. When the PER is less than 1%, the throughput drops slightly. According to PHY simulation results shown in Fig.6, if we set the PER threshold to $10^{-2}$, the selected transmission modes and the corresponding intervals of the ESNR can be defined in Table 4. The source nodes can therefore set the transmission mode according to this table and feedback of estimated ESNR.

**Table 4. Transmission mode vs. ESNR interval.**

| mode | 0 | 1 | 2 | 3 |
|---|---|---|---|---|
| ESNR interval | (-∞, -1] | (-1, 1.8] | (1.8, 4.8] | (4.8, 6.8] |
| mode | 4 | 5 | 6 | |
| ESNR interval | (6.8, 9] | (9, 13] | (13, +∞) | |

*Note: mode 0 means the ESNR is too small (the channel is harsh for communication) and the node should stop transmitting the packet.

Figure 14 gives the normalized throughput of different transmission modes. From it we observe that the lower mode has the higher normalized throughput. This is because the packet duration of the lower mode is longer, which leads to a higher PT-ratio.

In order to evaluate the performance of the proposed cross-layer solution, we compare the performance achieved by our cross-layer solution against that achieved by individual communication functionalities without considering the adaptive modulation and the cooperation between the PHY and MAC layers. We evaluate the network's performance over different channel conditions which can be equivalently described by the parameter ESNR. ESNR includes not only the noise but also multipath interference. Therefore in our simulation, we use different ESNRs to describe the varying channels. The ESNR ranges from -2dB to 15dB.

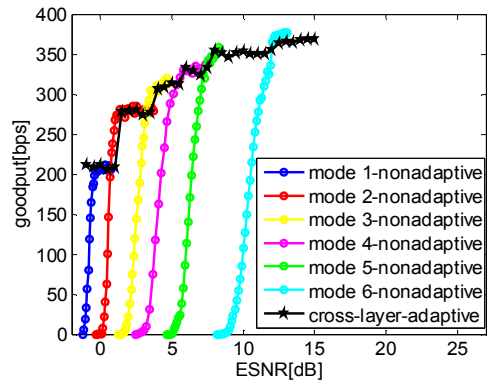

Figure 15. Goodput of different modes.

Figure 15 gives the goodput of the network with cross-layer solution compared with the traditional layered network with different transmission modes. From it we can see that, for each transmission mode, when the ESNR is smaller than what the mode requires, the goodput drops quickly. The goodput reaches zero when there is no packet that could be transmitted successfully. For the cross-layer solution, the sink node estimates the ESNR and feeds it back to the source node. The source node chooses the appropriate transmission mode according to the ESNR which keeps the goodput always achieving the maximum value. When the estimated ESNR is beyond the range that the receiver can successfully demodulate the packet, the source node will stop transmitting the packet for a while, thus avoiding energy waste.

## 5. CONCLUSION

In this paper, a new cross-layer design considering the interaction between the PHY layer and the MAC layer was proposed to optimize the overall system performance of the underwater acoustic networks. A modified CSMA/CA protocol was evaluated for its potential to increase the throughput for a common single-hop star topology. Modulation, convolutional coding and frequency diversity were jointly selected in an OFDM system to form a group of transmission modes to adapt to various channel conditions. The receiver was shown to be able to distinguish the packet loss reason caused by channel or collision, feeding back the information to the sender. According to the feedback information, the sender then logically modifies the ARQ/backoff and data rate to maximize the throughput. The simulation results reveal that the cross-layer solution effectively increases the throughput.

## 6. ACKNOWLEDGMENTS

This paper is supported in part by the China Scholarship Council (CSC) and by NSF CNS CRI award 1205725.